# Theories of Accounting: Evolution & Developments, Income-Determination and Diversities in Use

Angus O. Unegbu
Department of Business and Management Sciences,University of Kurdistan Hewler.
Email:unegbu4@yahoo.com

*The research is financed by University of Kurdistan Hewler but I hereby acknowledge the preliminary contributions of Late Dr L. A.Onojah.*

**Abstract**
Accounting frameworks follow stipulations of existing Accounting Theories. This exploratory research sets out to trace the evolution of accounting theories of Charge and Discharge Syndrome and the Corollary of Double Entry. Furthermore, it dives into the theories of Income Determination, garnishing it with areas of diversities in the use of Accounting Information while review of theories of recent growths and developments in Accounting are not left out. The method of research adopted is exploratory review of existing accounting literature. It is observed that the emergence of these theories exist to minimize fraud, errors, misappropriations and pilfering of Corporate assets. It is recommended that implementation prescriptions of these theories by International Financial Reporting Standard Committee and Practicing Accountants should be adhered to and simplified so as to avoid confusing and scandalous reporting of financial statements.

## 1. INTRODUCTION
The International Accounting Standards Board (IASB) was formed in 2001 as a successor to the former International Accounting Standards Committee (IASC), which was established to formulate and publish, in the public interest, International Accounting Standards (IAS) to be observed in the presentation of published financial statements and to promote their worldwide acceptance and observance (International Financial Reporting Standards - IFRS, 2007). International Accounting Standards Board (IASB) is responsible for establishing, monitoring and giving acceptable interpretations of the provisions of International Financial Reporting Standards (IFRSs). IFRS since inception has introduced numerous new useful, complex, confusing and/or expanding existing accounting frameworks. Frameworks of Accounting are raised from existing Accounting Theories.

Accounting theory is a material field in Accounting. Historically, accounting predates monetary economy. This was precisely, in the era of barter economy (i.e. exchange of goods for goods) when transactions were not only pre-determined by measurement but also by exchange values. The precept in which goods were exchanged at arms-length through concerted efforts of gathering, determining and measuring values are both pre and post-ante accounting. The Trade by barter period was characterized by measurement inequality, cumbersome in terms of production variety and coupled with the problem of coincidence of wants, were all-inherent in barter economy. However, the development of accounting theory was to ameliorate the inherent problems encountered in barter economy, unlike monetary economy. It is pertinent to understand the meaning, scope and application of a theory in humanities and management sciences in order to appreciate the work of accounting theory.

A theory according to American Institute of Certified Public Accountants (AICPA), (1970) is a structure that unifies the underlying logic or system of reasoning. Such theoretical structure, though abstracts from the complexities of the real world is designed to achieve a level of simplicity necessary for analysis. However, theory is useful in explaining, evaluating and predicting the phenomena associated with a given field of thought like in the case of accountancy. Osuala (2005), like Okoye (2003) views theory as an attempt at synthesizing, interacting and integrating empirical data for maximum clarification and unification. He added that every individual has a number of personal theories based on postulates and assumptions of varying degrees of adequacy and truth from which he makes deductions of various degrees of crucially and of course of accuracy. It will be useful to state that the word 'theory' is used at different levels even in the history of accounting. Accounting theory may mean purely speculative interpretations or empirical explanations of events for economic decisions. Accounting theory is defined as a cohesive set of conceptual, hypothetical and pragmatic proposition explaining and guiding the accountant's actions in identifying, measuring and communicating economic information to users of financial statement, (American Accounting Association (A.A.A). 1966). Wolk, Dodd and Rozycki (2008) opine that accounting theory consist of the basic assumptions, definitions, principles and concepts and how they are derived. They further assert that it includes the reporting of accounting and financial information. According to (Perara and Matthew, 1996), it is the logical reasoning in the form of broad principles that provide a general frame of reference to every accountant to evaluate and guide the development of new practices and procedures. It is the rationalization of the rules of accounting which further explains the manner in which accountants gather, record, classify, report and interpret financial data especially when monetary amount is determined in the financial statements. In the words of Hendrickson, (1992), accounting theory was defined as logical reasoning in the form of a set of broad principles that (1) Provide a general frame of reference by which accounting practice can be evaluated, and (2) guide the development of new practices and procedures. Accounting theory is used to explain existing practices and procedures to obtain a better understanding and to provide a coherent set of logical principles that form the general frame of reference for the evaluation and development of sound accounting practices. In accounting however, theory has loose and overlapping meaning with principles, concepts, conventions, doctrines, standards, rules, assumptions, tenets, postulates and procedures which are used interchangeably in this case. These doctrines however gave credence to the rational judgment, universal applicability, comparability, and acceptability of financial statements. Accounting conventions, unlike the laws of chemistry or natural science, are man-made-laws on data generation, recording, classifying and analyses of financial information that are at least in part of monetary character and interpreting the results therein for management decisions, Anao,(1996).

Unifying the views of American Accounting Association (A.A.A.) (1996), AICPA (1970) and Anao, (1996), accounting theory means a cohesive set of conceptual, hypothetical and pragmatic propositions explaining and guiding the accountants' actions in identifying, analyzing, measuring and communicating economic information to the users an informed decision. These principles represent the best possible guides based on reason, observation and experimentation. These rules are constantly changing, and hence resultantly influencing the business practices. These principles however, contradict and conflict the interest of statement users because various parties have different interests. Even though principles were developed from the opinions of the stakeholders (creditors, labour unions, management, accountants, teachers, auditors, journalists, financial institutions, government, tax authority, etc), their areas of diversities can hardly be resolved, Goldberg,(1949). As theories are evolving, some are either rejected or accepted or continually being revised or modified in order to keep pace with the increasing complexity of business operations and business risks. This is the nexus that empowers International Financial Reporting Standard (IFRS) its relevance. Accounting theory in recent time, has experienced tremendous growth and development, just like any system void of rules and regulations may encounter pre mature death and stagnation, barred from withstanding the test of time and may lack basis of evaluation and comparability, Macre,(1981). Globally known influential changer of many conceived and underling Accounting theories is International Financial Reporting Standard (IFRS).

The main objective of this study is to critically review the Origin, Growth and Development of accounting theories and their impacts on financial reporting. Other objectives are to explore accounting theory in resolving areas of diversities among users of financial statements. It further examines the various uses of accounting concepts and real income determination in the financial statements. Furthermore, to examine the extent to which accounting theory has influenced practices and development of accounting profession in recent times. It is obvious that the governments, financial institutions, professional and academic institutions and other users of financial report stand to benefit greatly from this research. The legislative, executive and judiciary arms of government also stand to benefit from the study in terms of policy formulation, administration and interpretations of financial statements for investment decision. The historical accounting periods in different regions of the world were critically reviewed and with particular attention to recent developments in accounting theory. The paper however reviewed the achievements made in accounting theory; and precisely in Europe, Asia, Athens, Mesopotamia, Great Britain and Africa. The review period is between $12^{th}$ - $21^{st}$ centuries. Emphasis was on evolution of accounting standards. The paper also offered possible suggestions for the improvement of accounting theory. The method used in gathering, recording and processing data is secondary source. The researcher traced the origin, growth and development in accounting by using different textbooks, magazines, journals and Internet services on accounting theory. The literature review has been organized into four main phases. First the evolution of accounting, followed by discuss on recent growth and development in accounting theory; secondly the fundamental theoretical accounting concepts; thirdly, the theories of Income Determination, and finally the area of diversities in the use of accounting information.

## 2. EVOLUTION OF ACCOUNTING AND RECENT DEVELOPMENTS IN ACCOUNTING.
In this section a review of evolution of accounting and its theories are explored. Recent growths and developments in the accounting arena are also discussed.

## 2.1 EVOLUTION OF ACCOUNTING

The early development of accounting system is traceable to the most ancient cities, in Mesopotamia, a home of number between 450 and 500 BC. (Keistar, 1965): Greece and Rome were cities where coinage was invented in about 630 BC (Chatfield, 1977) and China is where accounting systems were concerned with the recoding of merchants, temples, and estates (FU 1971). Keister (1965), further described the use of clay tablets impressed with the markings of the Cuneiform script by the Scribe, a forerunner of the present day accountant. The system though relatively simple by modern standards; the Mesopotamia economy did not require more advanced system to record its transactions and property among parties. Goldberg (1949) also recognized the recording of complex transactions of grain involving several individuals, a system of record-keeping (accounting) which is a clear demonstration that accounting is socially constructed. Chatfield (1977), saw the systems of estate records in part of Athenian Empire, by Zenon in terms of data collection, recording and analysis by several individual as responsibility accounting. This system employed by Zenon Papyri with respect to data generation, recording and analysis, (though elaborate and meticulous) were sufficient to detect error, fraud and inefficiency in the system. The Zenon Papyri approach had little concern for decision making, efficiency or profitability, and perhaps this feature might invalidate a lot of work that went into the operating system (Glautier, and Underdown 2001). The Zenon system was developed in the $5^{th}$ Century BC and later modified by the Romans. Goldberg (1949), saw the modification of Zenon in ancient Rome as the memorandum book (adversaria' in Greek) and the monthly transfer of entries to the ledgers ('codex tabulae' in Greek), from which today's ledger has derived its name 'codex'. This system of recording in ancient Greece and Rome according to Goldberg (1949) and Chatfield (1977), indicates that the accounting systems were mainly concerned with recording and exposing losses due to theft, fraud, inefficiency and corruption. It was not for decision making and assets protection. Gulman (1939), added that the accounting system at that time avoided financial reports to outsiders or determination of income or tax due to government and allied parties. The system still reveals that the accounting system at that period was of course fulfilling the societal needs and expectations of the users of financial statements.

Fu (1971) said the accounting systems that were mostly used by feudal and expansionist for merchants and estates in China, under Chou dynasty (1122 - 1256), allowed for large physical distances and several layers or hierarchies. Officials who were needed to collect taxes in the form of goods for use by the imperial government did so to ensure compliance. The surplus products however were collected for export and were used outside China, (Yameh, 1940). The system, though in details, covers several officials and large distances to ensure good administrative control through the appointment of higher-level officials as auditors who report at periodic intervals of ten days, thirty days and yearly as the case may be. The Chou system may presumably have stringent and appropriate penalties for non-compliance by defaulters, (Yameh, 1980). Ahmed (2000) argued that, funds accounting system exists in the form of general reserve fund, special reserve fund and reserve fund. The source of the goods, the purposes for which they were used, the frequency of taxes being levied and each tax ceiling were all bases of accounting system. Nwoko, (1990), in similar vein, observed that the earliest records known, which pre-dates monetary economy, were all accounting records, and were of ancient Middle Eastern Civilization of Egypt, Mesopotamia, Crete, and Mycenae. These were mainly records of physical quantities of goods.

Perara and Mathew (1966), opine that coinage was invented probably in Lydia at about 700BC as a result of difficulties experienced in maintaining the records and other inherent factors associated with barter system. The early accounting records were inscribed on stones and marble tablets in the Parthenon building accounts in Athens and Acropolis. Nwoko (1990) and Perara (1966) also observe that the Zenon Papyri which was discovered in 1915 contains information in business, agriculture, and construction projects of the private estate of Apollonius kept under the accounting system. These records were kept in surprising and elaborate system that had been in Greece since the fifth century BC. The Zenon accounting system had provisions for responsibility accounting; written records of all transactions, personal account for wages paid to employees, inventory records, and records for assets acquisitions and disposals. In addition, it contains evidence of auditing of all accounts, (American Institute of Certified Public Accountants (AICPA), (2006).

### 2.1.1 THE CHARGE AND DISCHARGE SYNDROME

The early Greeks and Roman accounts were kept under "charge" and "discharge" principle, comparable to modern day receipt and payment account, (James 1955). The medieval system of record keeping used in England during the middle ages had many features of ancient accounting system and remained in use until the nineteenth century as "charge" and "discharge" (James 1955 and Nwoko 1990). James' perception of 'charge' and 'discharge' which was similar to the present day receipt and payment account or cash book was presented in the form of:

**Charge and Discharge account**

| Arrears | $ | Expenses | $ |
|---|---|---|---|
| Rent and farms | x | Money delivered (Total discharge) | x |
| Other receipts | x | The balance (remainder) | x |
| Total (charge) | x | Total (discharge) | x |

Source: Nwoko, 1990

The feudal socio-economic system requires that surplus be generated but does not have any perceived need to measure the efficiency by which the surplus was generated. Moreover, no notion of income or return on capital employed was in practice at that time. The manorial system dwells primarily on the interlocking check on the honesty of different levels of officials in a stratified and regulated society, (American Accounting Association, 1964). The charge and discharge syndrome was surprisingly durable, lasting from twelfth to nineteenth centuries, (James 1955). The book keeping for merchants was however in single entry form, rather than by charge, prior to the arrival of an Italian Monk-Luca Pacioli in England, the acclaimed father of double entry bookkeeping system, (Litleton, 1966).

### 2.1.2 THE COROLLARY OF DOUBLE ENTRY
The Normans imported the charge and discharge book keeping system, originated in the Mediterranean zone into Europe. It was in Europe that the next significant development in accounting emerged (Nwoko, 1990). It occurred in Italy, between thirteenth and fourteenth centuries, probably because of single entry system which was inadequate to ensure effective internal control system, income determination, security of assets, employees' contribution to profit, and separation of private property from business. The single entry system of recording did not withstand the changes in size and nature of business organizations including the methods of providing for depreciation, (Rorem, 1937). The double entry system sensitized merchants to distinguish between positive (+) and negative (-) entries or increases in assets and decreases in liabilities, (Paul, 1985). Nwoko (1990) emphasizes that those positive entries that increased assets or reduced liabilities are: cash receipts, sales to customers, payment to creditors, discount received. While negative entries that increase, liabilities and reduces assets are cash payment, purchases, discount allowed, and payment by debtors. The Latin words Dare (to give) and Avere (to receive) were given in English as Credit (Cr) and Debit (Dr) respectively, and were employed only on the completion of the venture. The balanced books of accounts however dates back to 1340 and were produced by the Massari or Stewards of the commune of Genoa, before it's widespread use in Italy and beyond in 1400, (Pyle, et al 1980).

The worldwide use of double entry however owes a lot to the work of an Italian Monk, and a Franciscan friar, in 1494, to Luca Pacioli. Pacioli's first printed work or treatise was on algebra, titled: **"Summa de Arithmatica, Geometrica, Proportioni et proportionalita (i.e. everything about Mathematics, geometry, and proportions)".** It was developed to ensure that every transaction has equal and opposite reaction, (Mike & Fred, 1983). The treatise contained a section on book keeping entitled **"De computis or Scripturis (i.e. computations and records),** which was separately published in 1504 and translated into many languages. Pacioli, however did not lay claims as the originator of double entry as he was only describing what Italian Merchants were using for over 200 years, (Paton and Littleton, 1940).

Nwoko (1990) recognizes Grammateus of Schreiber, (1518), as mathematician of no mean repute who wrote a book on algebra and bookkeeping. Jerome Cardam, an astrologer, physician, scientist, mathematician and professor of medicine, like Simon Stevin, a Dutch Mathematician with various claims to fame, wrote also on double entry bookkeeping in 1602 Institute of Chartered Accountants in England and Wales (ICAEW, 1975). The new concept however was only describing a system in practice which lack general rules and principles and did not however show clear method of calculating profit. It further revealed that provision for depreciation was virtually absent including method of drawing up a balance sheet, (Edey, 1970). Perera and Mathews (1996), had a strong view that the initial development of double entry bookkeeping in the Italian city states experienced long period of stagnation, probably because of its non-acceptance in Europe: England, Germany, France and in Italy. Commercial activities at these periods were inactive, though due to size and type of business, which also encouraged the use of single and double entry bookkeeping, without regular closing of entries and income determination (Baxter, 1981). The socio-economic changes from "a green land" (craft techniques) to "one dark satanic mills" (factory production), were powered by machinery, building of factories and towns, separation of ownership and control (capitalism), emergence of large-scale industrial and commercial activities and accounting system in Europe, precisely in England (Pyle, et al 1980). The side effects of these changes on accounting were profound in the development of recording, measuring and disclosure requirements in factories, railways and aggregation of labour and capital equipment. New system of production, ownership and control of assets including methods of providing for depreciable assets were based on accounting assumptions, (Ola, 1985). Charge and discharge system could not meet the aggregation of capital equipment, calculation of product costs, inventory valuation, income determination and depreciation of factory fixed assets; to the development of new accounting system that could replace charge and discharge system in a new changing technological environment (i.e. depreciation accounting, (Jennings, 1990).

Chafield, (1977) like Omolehinwa (2004) observes that where depreciation was not charged, costs were understated, profit overstated and dividends were paid out of capital. Prior to company taxation, the early theories of depreciation including replacement cost accounting contend that there was no need for depreciation if the assets were maintained in good condition. This theory however, would produce as many problems as it was meant to solve, (Paul, 1985) Stoner, et al, (2002), did observe also that the emergence of labour in factories led to the need for the development of systems on how to pay wages, overtime, bonuses, piecework and as well as managing the large number of employees that were necessary for the new industries. Akanni (1998), observes that in every organization, both the employers and employees have sworn to be enemies, though one cannot do without the other because employees need wages and employers need labour for production. Accounting systems for wage and production must be designed for that purpose. The economic and legal changes resulting from industrial revolution, particularly to the aggregation of capital, labour and company legislation brought pressure on the accounting systems that would put the various parties at par. Accounting system that could address aggregation of capital, methods of labour remunerations, depreciable assets, production cost, and income determination was developed, (Dopuch and Sunder, 1980). Similarly, Golberg (1949), saw companies and bankruptcy legislation in UK, dated to 'Bubble" Act 1719, as part of industrial revolution which declared unincorporated companies not formed by Royal Charter as common nuisance.

According to Mike and Fred, (1983), the practice in accounting could not develop properly to meet the ever increasing demands of a complex society without reference to a coherent underlying theory for true income determination. Bierman and Drebin (1972), in a similar way observed that there was no response to the need for a consensus on the nature and application of Generally Accepted Accounting Principles (GAAP) in given nation. The Institute of Chartered Accountants in England and Wales (GAAP) has since in 1942 been publishing recommendations on Accounting Principles. The development of a framework of principles to underpin the preparation of financial accounts, perhaps received greater attention by US government in 1934, as a result of great economic depression. This act however, led to the establishment of Security and Exchange Commission (SEC) in 1934 to prescribe the principles to be applied by the American Accounting Professionals in the preparation of financial statements. The American Institute of Certified Public Accountants (AICPA), took more positive approach in the establishment of a coherent set of accounting principles. The AICPA established the committee on Accounting Procedure (CAP) in 1938 which was replaced in 1959 by the Accounting Principles Boards (APB). In 1973, this in turn was replaced by Financial Accounting Standards Board (FASB). The need for international accounting principles that would cope with the need and increasing multi-national nature of business, led to the establishment of International Accounting Standards Committee (IASC) in July, 1973. Chambers (1966), observes that Generally Accepted Accounting principles (GAAP) has over 100 lists different from about 70 lists by Spacek put together in all about 170 GAAP exist and now merged to 13. He further stressed that accounting principles can also be described as concepts; conventions; postulates; standards; doctrine; tenets; assumptions; rules and regulations governing the preparation and presentation of financial statements. Bierman and Drebin (1972) were more pragmatic in grouping accounting principles into three. First is the "assumptions about the world", second, the "operating conventions" and third is the "quality considerations" for the purposes of true income determination in the financial statements. The three groupings according to them would however, unify the interests of the various parties, all things being equal. The ideological thoughts of Bierman and Drebin (1972), is structured under the Generally Accepted Accounting Principles (GAAP) in a block diagram:

**GAAP IN BLOCK DIAGRAM**

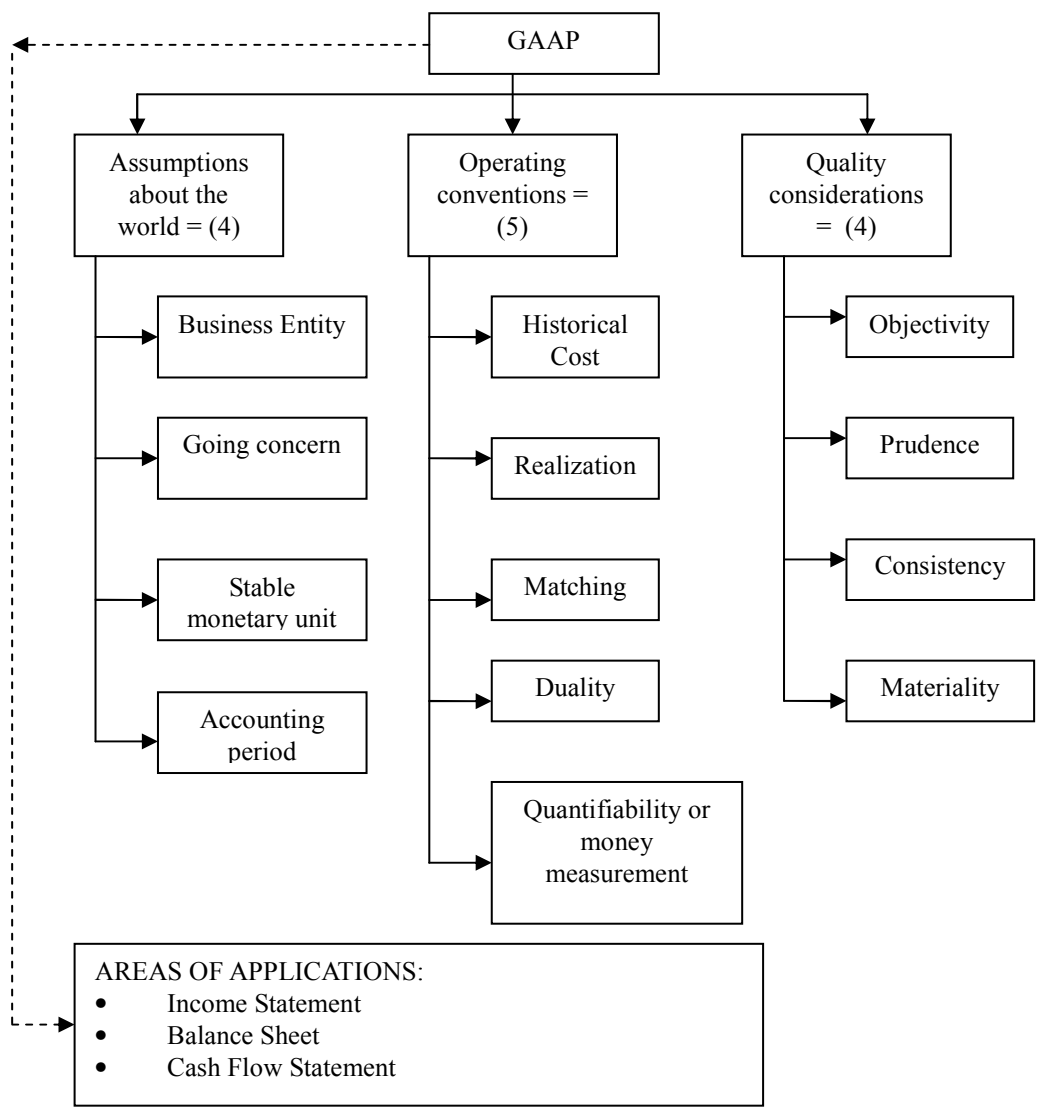

Source: Hendrikson, 1992

## 2.2. RECENT GROWTHS AND DEVELOPMENTS IN ACCOUNTING
Accounting in recent years, has made significant impact on socio-economic and political development especially on recording, preparing, interpretation, auditing and management and investment. Other impacts include merger, acquisition, planning, controlling, and storage of business operation. Above all, is the impact on the decision making process, (Remi, 2006).

### 2.2.1 Regional Grouping of Associations
Glautier and underdown (2001), observed that regional grouping of accounting associations indeed developed accounting principles, peculiar to their culture, religion, government policies, political and socio-economic environment. These were later absorbed into International Accounting Standards Committee work plan. This approach assured determination and comparability of profit, revenue, expenses, net assets, and liability internationally. Accounting bodies now regulate and ensure compliance to the application of GAAP, (Adeniyi, 2004).

### 2.2.2 Consultative Committee of Accounting Bodies (CCAB)
The institutional structure of accounting profession made possible the formation and amalgamation of various accounting bodies in 1965. By 1970, ICAEW formed an Accounting Standard Steering Committee (ASSC) now called Accounting Standards Committee (ASC). Early Committee members include; Institute of Chartered

Accountants of Ireland (ICAI: 1970), Association of Chartered Certified Accountant, (ACCA, 1973) Institute of management Accountants, (ICMA, 1976) and the Chartered Institute of Public Finance and Accountancy (CIPFA: 1976). Others include Financial Accounting Standard Board (FASB); European Economic Community (EEC) now European Union (EU) Security and Exchange Commission (SEC), Financial Reporting Standard Board (FRSB). They were saddled with the responsibility of reviewing standards on Financial Accounting and reporting and to publish consultative documents on maintaining and advancing accounting standards. Also to propose to the councils the best statements of standard accounting practice. Consultation was usually made with representatives of finance, commerce, industry, government and other persons concerned with financial reporting. This however, resulted to uniform accounting standards and practice all over the world, (Nwoko 1990).

### 2.2.3 The Use of Exposure Draft (ED) and Letter of Intent (LOI)

Justification and application of (ED) and LOI said Mootze (1970) is to enable various professional associations and users of financial statements all over the world to first analyze the accounting implication and adopt a uniform position before the publication of Generally Accepted Accounting Principle (GAAP). This however, will encourage uniformity, comparability and convertibility of financial statement in different currencies across national boundaries, -Robert, (1999) and Adeniyi (2004).

### 2.2.4 Setting Accounting Standards

Postulates, assumptions, tenets, principles, rules, laws, and theories said Mootze, (1970), constitute the basis of practicing accounting. Violation of GAAP may result in qualifying the financial reports. Treatment of incomes, expenses, assets and liabilities, should adhere to the normal accounting standards. Otherwise there will be no basis of truth and fairness in the financial report (Robert, 1999). The concept of double entry or accounting equation (A=C+L) shows why trial balance or balance sheet must always balance, (Nwoko 1990).

### 2.2.5 Training, Workshop and Seminars

Institutions of higher learning in different regions have adopted the training programmes and researches for the development and improvement of accounting standards. This process is to ensure uniformity in the treatment of business transactions, (Stoner et al, 2002). Workshops and seminars are being organized in different regions by accounting bodies including governmental and non-governmental organizations. The objective is to enlighten, educate and inform users on how to prepare credible financial statements, especially on transparency and public accountability. Nwoko (1988) did observe that continuous training and development have made great impact on public cost consciousness and accountability.

### 2.2.6 Information and Communication Technology (ICT)

Millichamp (1990) views accounting information as data processed from source documents. The source documents usually include receipts, vouchers, invoices, Local purchase order (LPO), cheques stubs and books of original entry which must be tested for proof of arithmetical accuracy (trial balance) for the preparation of final accounts, (trading and profit and loss account, balance sheet). Other information that are statutorily communicated to the users are": note to the accounts, auditors report, cash flow statement, value added statement and group accounts, produced either in hard copy or by electronic device. The Internet service or on-line system has made accounting reports to be produced and communicated on time to users with high degree of accuracy (Robert: et al 1990).

### 2.2.1 Audit Ordinance or Guidelines

The Company Act 1948, and 1968 as amended, the constitutions of different countries such as;  
US, UK, Italy and the constitution of the Federal Republic of Nigeria 1979, 1989 and 1999 as amended stipulate the use of public funds. The Finance Control and Management Act 1958, the Audit Ordinance Laws (1956), established the Consolidated Revenue Fund (CRF), Development Fund (DF) and the Contingencies Funds (CF) to ensure proper control of public funds. The systems state the basis of government accounting and audit. The Audit Ordinance (1956), companies and Allied matters Act (CAMA) 1990, outlined duties, responsibilities, appointment, tenure, removal and retirement of Auditor General for the Federation. The relevant laws and edits also respectively govern the state and local governments.

### 2.2.2 Discipline and Sanctions

Accounting institutions, associations and government have absolute control on compliance and adherence to financial regulations, treatment of business transactions, and code of conduct, preparation and presentation of financial statements, (Pyle et al, 1980). Discipline and sanctions await erring members for non-compliance. Nevertheless discipline is seldom carried out for lack of hard evidence. The presence of Economic and Financial

Crime Commission (EFCC) and Independent and Corrupt Practice Commission (ICPC) are now tools against corruption in Nigeria economy.

**2.2.3 Researchers**
Universities are research geared, research institutes and accounting associations now pursue with vigor by way of modification and update on accounting standards in the treatment of business transactions. Accounting teachers have offered ideas and suggestions in the formulation and development of accounting theories. Accounting software package are now available especially on inventory valuation, ledgers, income statements, balance sheet, cash flow statement and budgeting. (Robert et al (1990).

**2.2.4 Data Processing and Information Technology (DPIT)**
Accounting however predates computer. Their integration is inseparable on ground of accuracy, quality, timeliness, speed, and storage. Akanni (1998) like Robert et al (1990) configured computer device for business and accounting operations as presented in Diagram 2 below:

**DATA PROCESSING IN BUSINESS**

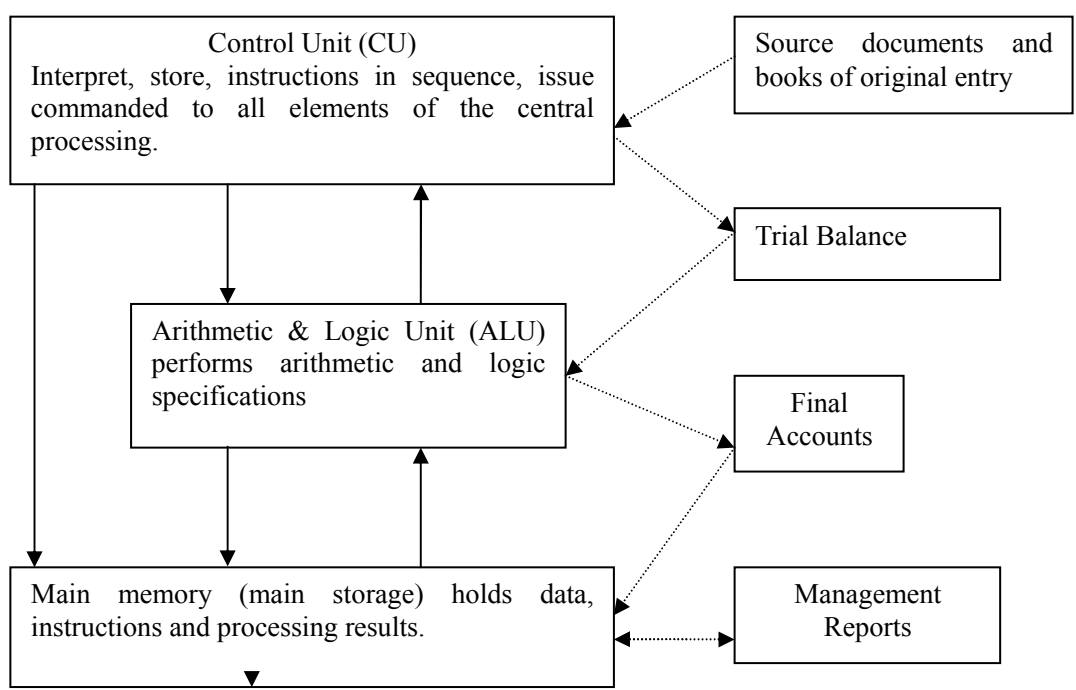

**Source:** Akanni 1998 (modified) p. 241

The explanation of the above figures shows that the infusion of accounting and computer is however further strengthened by the methods and stages of data gathering, recoding, analyzing, summarizing, communicating and interpreting the financial statements. These stages are also referred to as data processing in business, Robert et al (1999). The strong tie that exist between accounting system and electronic data processing system as configured above lies on the sequence of:
  i. Data generation from the source documents.
  ii. Issuance of instructions and commands to the various processing units or elements by the (CU).
  iii. Performs arithmetic and logic operations on the source document by (ALU).
  iv. Test of the books of original entry by means of a trial balance for mathematical accuracy.
  v. Preparation of final reports for management and others for efficient planning and control.
  vi. Store information and instructions in the storage units or main memory and auxiliary storage unit Robert et al (1999).

**3. Review of Relevant Theories of Income Determination**
In preparation of Income Statements, there are underlining theories that are material in the structuring and determination of what is considered as Income. The theories of real income can be considered under the following four perspectives:

3.1 Business Entity Assumptions (BEA): Husband (1954), contends that business can be separated from its owners and the environment in which it operates is necessary in order to set a boundary to the accounts. Only the transactions directly affecting the entity are recorded in financial statement. Omolehinwa (2003) also observed that the separate legal personality is assumed as business has a right to acquire assets and incurs liabilities as distinct from its owners. The business has right to sue and be sued like any other person. Both agreed that it can sometimes be somewhat arbitrary, particularly for small and medium enterprises where the affairs of the owners and the businesses are often inextricably interwoven. This process would however give rise to distortion in real income determination, especially where information are not readily available about private expenses of the owners as distinct from the firms. Early advice and proper accounting records will however eliminate the pending danger of not separating private expenses from business expenses. However, what Husband (1954) and Omolehinwa (2003) did not recognized was the inability of the court to imprison the entity as individual can be sentenced and imprisoned, except those who acting in that capacity.

3.2 Going Concern Assumption (GCA) means that in drawing up financial statements; the entity will continue to exist in its present form into the indefinite future (perpetuity) (Fremgen, 1968). It is further stresses by (Fremgen, 1986) that organization will continue to exist for life as far as the firm can meet its immediate and long term financial obligations. GCA, however ensure that assets should also be valued based on their economic useful life, cost, degree of usage and residual value for purpose of real income determination. The controversy to going concern assumption is that a firm could be compelled to go into liquidation if it cannot meet its short term and long term financial obligations as they fall due. Other reasons for a firm's liquidation may include:

i . On litigation by creditors or by an order of the court.
ii. Proclamation by the government
iii. When the firm operates outsides its memorandum and articles of associations. (ultra vires)
iv. When the activities of the firm are illegal, injurious to health and against the public policy
v. On mutual agreement by the owner and stake holders.
vi. On completion of a particular venture or contract and;
vii. By an act of God (disasters, death etc)

Omolehinwa (2004) added that in valuation of assets, firms may use different methods which may give different values in the financial statement. These methods include first- in- first- out, weighted average, highest- in- first-out and next -in- first- out. Depreciation also has different methods of valuing assets. These methods are straight line method, reducing balance method, sum of the year's digit method, revaluation method, annuity method, sinking fund method and production units or hourly method. The choice criterion among the various methods of assets valuation could result to over or understated profit in the financial statements. The going concern assumption can however be realistic to some extent if the firm will be able to meet its routine and long term financial obligations as they fall due. Otherwise, organization may abort this assumption by going into liquidation. The strength and weakness of a firm will possibly predict or signal to interested parties on pending danger for immediate action by identifying the various strengths, weaknesses, opportunities, and threats (SWOT) of a firm.

3.3 **Stable Monetary Unit Assumption (SMUA):** states that the value of the monetary unit used in drawing up the financial statements is constant over time. The validity of this assumption is somewhat questionable (Fisher, 1980). Its existence is apparently based on the additive nature of accounting data. Within a set of accounts all the numbers that are capable of addition and subtraction exist. It is clear that even among infant school arithmetic, it is not possible to add or subtract unlike terms or items and get a meaningful result like two elephants, five apples, and ten motor vans do not make a meaningful quantity of seventeen (Fisher 1980). The assumption however is that money values of property, plant and machinery, stock, debtors and cash obtained at different times are summed up in a balance sheet; which are identical, even though the transactions may have taken place at different times. The stability assumption also overlooks the purchasing power of money which is constantly changing due to inflation and other factors, (Buckmaster, and Brooks, 1974). This assumption also negates the time value of money, as it erroneously, compares N200.00 profit in 1980 with N200.00 profit in 20013, especially when interpreting the financial statement over two decades. It is however, glaring to note that the value of a firm's profit for different accounting periods cannot claim to be the same because of the presence of inflation or changes in purchasing power or price level changes. Generally, the rate of inflation can be used as yardstick to determine the appropriate rate for cost of capital. The use of appropriate accounting index could restate and resolve the differences between the accounts prepared under Historical Cost Accounting (HCA), with the Current Purchasing Power Accounting (CPPA).

**3.4 Accounting Period or Periodicity Assumption (PA)** believes that the continuum of time can be subdivided into a number of discrete time periods (accounting periods) (British Institute of Management Information (BIMI, 1984). According to Keynes (1980), in the long run we are all "dead", implies that the net income of a firm prepared under this assumption (discrete time period), may not after all be realistic because of the presence of inflation and changes in price level. He asserts that the net profit usually has the components of unrealized profit or uncollectible realized revenue that may eventually go bad in the form of bad debts. The provision for losses may however not cover the total debtors during the accounting period. Similarly, the fundamental problem, be it in the short or long run, is the method of stock valuation. Stock valuation could be based on full cost or marginal cost, straight-line or declining method, especially when the firm is operating either at full-capacity or under-capacity. Where there is under-capacity utilization the absorption rate is likely to be high if the depreciable amount is always high, it reduces the profit figure in the financial statements, vice-versa.

**3.5 Operating conventions are classified into five as indicated below:**

**3.5.1** *Historical cost Convention (HCC):* Forms the basis of valuation used in the preparation of published financial statements. That is, all assets are shown in the accounts at the cost of acquisition. The word 'cost', said Horngren and George, (1990), is intricate, complex and confusing. This is because cost may mean different things to different people at different time, place and event. They assert that cost is intricate when referred to as expired or futuristic, production cost or period cost, direct or indirect cost. It is complex when referred to as variable cost, fixed cost, semi variable cost, semi fixed cost, marginal cost, absorption cost, sunk cost, conversion cost and opportunity cost. Confusion may arise when certain cost attributes are not or are to be capitalized for purpose of assets valuation. They added that if asset was acquired under a given scenario, two different accountants may arrive at different value judgment about the asset as follows: (Pyle et al, 1988).

|                       | $   |
|-----------------------|-----|
| Purchase cost         | x   |
| Agreement fee         | x   |
| Installation charges  | x   |
| Improvement cost      | x   |
| Development cost      | x   |
| Total cost            | x   |

Since assets are valued at historical cost, for purpose of income determination, the confusion may arise if purchase cost or part therein or total cost is regarded as historical cost. The net book value arrived at as a result of deducting depreciation charges based on different depreciation methods may be different, hence the difference in net profit, (Okoye, 1997).

**3.5.2** *Realization Convention (RC):* It suggests that it is when contractual relationship between the buyer and the seller was completed that the amount of revenue is recognized and recorded in the books of accounts but not necessarily when cash is collected. Turpin and Stein, (1986) assert that accounting period imposes serious problem to realization principle as the entity needs to have recognized the transactions during the accounting period but not when contractual relationship was completed. The problem was further stressed by American Accounting Association (1965) as thus: an entity produced motor van in year one for $30,000, stores it through in second year, sells it in third year for $60,000 and collected cash in the fourth year. Realization convention argued that the revenue and profit element of $60,000 and $30,000 respectively are realized in the third year when the contract of sale was completed, but not in the accounting period - year one. Confusion may arise when revenue is realized on production basis rather than when goods may have been sold or service discharged, especially as in the case of government contracts. Sometimes, contracts may last over and above the accounting period and revenue may be recognized on the basis of the extent of work done or production, Hylton, (1965).

**3.5.3  Matching Convention (MC_):** Leads to the matching of entity's revenue generated with the expenses incurred. The matching process also deals with the allocation of capital costs between periods. The combination of the matching principles with that of realization gives rise to the accruals system of accounting, Hylton (1965). This means that profit will be recorded at the point of sales, whether cash is received or not. Similarly, the matching principle leads to the association of an expense with the revenue that it generated, irrespective of cash payment connected with the expense. Profit or net income is however reported in the financial statements, when costs consumed or incurred during the accounting period are matched with revenue realized either on cash or accrual basis during the same accounting period.

**3.5.4  Duality Convention (DC_):** This is associated with the system of Double Entry Book-keeping invented by Luca Pacioli in 1494. It requires that every accountant enters both aspects of every transaction in the books of account because every action has an equal and opposite reaction, (Sterling: 1972). For every entry in a ledger an entry of equal magnitude must be made on the opposite side of the ledger or another ledger. Thus, the sum, of the entries on both sides of all ledgers, barring errors, must always be equal. Hence, the arithmetical accuracy of all ledgers is further tested through the agreement of a trial balance for absolute proof. The trial balance and profit of arithmetical accuracy of all the postings in the ledger however assert that my agreement is not an absolution proof because of errors of: (1) Omission (2) Original entry (3) Commission (4) Compensation, (5) Principles, (6) Additive and (7) Transposition, Omolehinwa (2004).

*3.5.6* **Quantifiability or Money Measurement Convention (MMC):** Says that all items included in the accounts must be measurable or quantifiable in terms of monetary unit. This assumption suggests that it would be insufficient to include non-quantifiable items in traditional accounts merely using an ordinal ranking for them, (Bierman, 1972). Placing or fixing monetary value to an item may somehow be confusing because of the presence of inflation or changes in price level value judgment however affect materiality concept for money measurement, as there is no uniformity relating to perceived value of an item in the book of accounts as value may vary from person to person and from firm to firm. (James, 1955).

*3.5.7* **Quality Considerations (QC):** Accountants will however strive to achieve certain qualitative characteristics in the application of the operating conventions when preparing financial statements as shown below:

*3.5.7.1* **Objectivity** states that entries made in the ledger shows that accounts must be capable of verification by an independent party. This would ensure that financial statements are free from bias and minimizes the possibility of subjective judgment by accountant (ACCA study pack, 1988). The issue of objectivity test is relative as IASC (2001), observed that even historical cost account cannot be completely objective. The contrast of historic cost accounting system lies on both valuation techniques and method of estimating assets life span, scrap value and acquisition cost. This implies that different accountants may have different values for the same assets, and different depreciable values, thereby resulting in increase or decrease in net profit. The verifiability of all entries in the ledger and accounts, however show how accountants can examine, verify, detect and prevent frauds, errors, embezzlement, misappropriation, pilfering and corruption.

*3.5.7.2* **Prudence or conservatism (C)** Connotes that where an accountant could deal with an item in more than one way, his choice between the alternatives should give precedence to which provides the most conservatism result, (Moore, 1972). The principle also states that in stock valuation: if the current price is lower than the cost of acquisition, the stock should be recorded at lower of cost or current price. And where assets have appreciated in value, the gain should not be recorded in the books of account until the assets are sold. Conventionally, the principle further states that accountants should anticipate for no profit but make possible provision for all losses, (Chambers, 1966).

*3.5.7.3* **Consistency consideration (CC).** This states that where a transaction or economic event is repeated in different time periods, then the accounting representation should be the same in all time periods. The consideration however does not preclude mistakes being rectified nor accounting treatment being altered when the changes are beneficial in terms of giving a better representation of the economic reality, (Burk, 1973). Hendrickson, (1992), however, stresses that where the accounting treatment is changed from straight line to a declining balance method of depreciation, the effects of the change and the position under both the original and revised accounting treatment should be clearly shown. The consistency consideration forms the basis of uniformity and comparability of financial statements within and outside the accounting periods, especially on target profit.

*3.5.7.4* **Materiality.** It states that the way an item is treated in the accounts should depend upon its magnitude. To classify an item as material depends upon the influence the item will have on the interpretation of the amounts SSAP 4. The accounting treatment of government grants state that the amount of the differed credit should, if material, is shown separately in the balance sheet. Macre (1981) like Nwoko, (1990) viewed 'materiality' as being subjective, as what may be material to one entity may not be material to another entity. Both agreed that for an item, to be material, it must be relevant, the value can be spread over and above one accounting period, the size or magnitude of the item could lead to distortion in the financial statements, and its inclusion or exclusion from the financial statements will be misleading. Chase (1979) adds that the significance nature of an item, will however determine its materiality effect in the financial statements.

## 4 AREAS OF DIVERSITIES IN THE USE OF ACCOUNTING INFORMATION

No system in the world may lay claims to be absolutely watertight. Accounting as a system is not an exception (Robert, et al, 1990). Though accounting grows homogenously with socio-economic and political development, it however encountered some endemic perils, which similar professions indeed encountered. Some of these perils include:

### 4.1 Cultural Diversity
Stoner, et. al (2002), observed that culture is an implicit factor to the organizational and work force development. The integration of different cultures and ideological taught to form a common force in establishing international accounting standards worldwide took centuries to materialize. Today the situation is however not too different as cultural imperialism on developing nations by the advanced nations had effects on early accounting development. Some religions especially Islamic religion does not encourage interest payment on loans and advances which however negates business ethics of capital growth and appreciation or increase in business income, (Robert et al 1990).

### 4.2     Language Barrier or Linguistic logy
The major drawback in early accounting development was the medium of expression and communication apparatus. Language barriers, among nations prevented early good intentions to stimulate and form a body charged with the responsibilities of formulating accounting principles, training and research. The method of communication was crude and mostly by letters which were virtually absent in developing countries (Akanni, 1988).

### 4.3     Early practitioners were not research oriented.
The early accounting practitioners were not research orientated. And those who lay claims on research did not find it easy after all as materials and sources of inspiration were difficult to obtain, (Benjamin, 1990). Generally, their level of education, and the so-called accountants at that time had no broad accounting knowledge. Even in the 21$^{st}$ century, most people especially in the banking, insurance and public sector organizations were often called accountants, even when they have no accounting background. This indeed limits their research capability and ability to formulate accounting principles Pyle, et al, 1980).The situation at that time suggests that workers are judged by their ability and not by their disability

### 4.4    Disagreements between academicians and practitioners
The superiority complex among practitioners and academic accountants did not timely restore peace and unity to encourage early development and growth in the accounting profession. The superiority complex according to Anao (1996), has not been entirely written off even in the 21$^{st}$ century. The two bodies that should have integrated their ideas and focus on developing the best accounting practice, have no early unified objectives.

### 4.5     Diversity and Complexity in government policies
The status-quo of independence immunity of sovereign nations determines the nature, scope and application of accounting systems peculiar and suitable for its socio-economic and political environment, (Glautier, and underdown 2001). It is however possible that those sovereign nations in line with their cultures, do formulate, design, and implement accounting systems which have less reference to other nations and such accounting systems will be void of international comparison. This is further aggravated by instability in monetary units and changes in socio-political economy, (Mootze 1970).

### 4.6    Judicial Application and Interpretation of Business Transactions
Environmental factors, culture, religion and political power had a great influence on the interpretations and treatment of business transactions among nations. Prior to a decided case by Justice Joyce in 1904 in Garner V. Murray, solvent partners share deficiency in capital in proportion to their profit sharing ratio. However Justice Joyce inversed the old rules in favour of the proportion of capital contribution by solvent partners. He argued that capital loss does not excavate from ordinary business operations, but from equity, (Paul, 1985).

### 4.7     Money as a unit of measure
According to Millicham (1990), translation of foreign subsidiaries (with different monetary unit into parent company which is based on exchange rate estimates and the magnitude of inflation over time, makes the international comparability of financial statements between two accounting periods very difficult. However in accounting, money is the only unit of measure, and naira amount exchange for dollar or pound sterling may not have the same value over time. The instability in exchange rate or monetary unit is however more common in developing countries. Even within the same country, the monetary unit remains unstable.

### 4.8    Non-Unification of Various Ideological taught by various Bodies
Nwoko (1988) observed that accounting bodies formed in different countries with different or similar objectives have been influenced by their indigenous cultures, socio-economic needs, political environments, incessant change in government policies and programmes. An attempt to unify these ideological thought into a united whole could result in a Herculean task to publish a complete document on accounting theory, (American

Research Journal of Finance and Accounting
ISSN 2222-1697 (Paper) ISSN 2222-2847 (Online)
Vol.5, No.19, 2014

www.iiste.org

Accounting Association (AAA, 1936)

### 4.9 Illiteracy
Before Lucas Pacioli, published his first book on double entry in 1494 there was hardly a system designed specifically for training and research in accounting. This period of "Dark Age" culminated into absolute ignorance and lack of statistical data towards early accounting development. Today, the situation is however different only to the extent to which we appreciate the presence of science and information technology, (Mootze, 1983).

## 5. SUMMARY, CONCLUSION AND RECOMMENDATION
Under this section a summary is made and conclusions drawn, cumulating into specific recommendation.

### 5.1 Summary
Accounting unconsciously developed from socio-economic and political needs of the society by tracking down the historical and current events in business and economics. The inherent problems of measurement, proportion, recording and coincidence of wants eased out by the introduction of standard unit of measurement. The growth in business that culminated into industrial revolution compelled accounting to move to another stage of development called decomputis, or 'charge' and 'discharge' system of bookkeeping. This system however did not facilitate the determination of profit because it lacks method of inventory valuation, cost ascertainment and provision for depreciation. The emergence of double entry system was to minimize fraud, errors, misappropriation and pilfering of assets. The system in most cases allowed equity owners to have confidence on the works and reports of the stewards (management), who were entrusted with the capital assets of the owners. The subsequent issues and development in accounting relates to the Generally Accepted Accounting principles (GAAP), a period when owners entrust their resources to the management group for target objective. Auditing and investigation however emerged to resolve conflict among users of financial statements. Users however, with the exception of management, gain assurance on the financial statement when auditors certify that the accounts have been prepared in line with the generally accepted accounting principles. Decisions by stakeholders on investment, takeover, merger and acquisition were normally based on non-qualification of auditors' reports. Finally today, accounting packages cum information technology and computing are readily available to ensure timely production of financial reports at minimum cost, high speed and accuracy.

### 5.2. Conclusion and Recommendation
Accounting like business and economics or any other system has experienced changes, modifications, updates and improvement in recent years. Stagnation between 1400 to early 1990 was due to cultural, political and ideological differences, government policies, and language and currency barriers. Others include lack of statistical data, non-availability of research personnel and institutions, illiteracy and superiority complex among academic and professional accountants. The situation is however, better off now than before because of the introduction of regional grouping, international accounting standard committee, exposure draft and statement of intent, including the availability of research institutes. These developments have made possible the universality and comparability of financial statements regionally and internationally. The current pressures exacted on contemporary accounting decisions were unresolved issues in accounting history. Individual interests, place, time, and event, have significant influence in computing cost, revenue, expenses and even choice of depreciation. Significantly, costs may mean different things to different people at different time and place. There are different methods of providing for depreciation in which the choice will however have effect on the net profit. The conflict resolution on current pressures exerted on contemporary accounting decisions includes consistency and absolute adherence to the prescribed accounting standards and financial regulations. The pressure exerted on contemporary accounting decisions is unified under a general pattern in which all financial records and reports are presented and adopted. Audited financial reports, similarly gave strength to universal acceptability and less biased though on the current pressures exerted on contemporary accounting decision.

It is recommended that implementation prescriptions of these theories by International Financial Reporting Standard Committee and Practicing Accountants should be adhered to and simplified so as to avoid confusing and scandalous reporting of financial statements.

**References**
ACA (2003), Foundation Accounting Framework, Paper 1, Lagos Wyse Publications.
ACCA Study Pack (1988), Level 2 Paper 214, Cost and management accounting II, London, Financial Training Courses Ltd.
Adeniyi A.A. (2004), Management Accounting Edition, Lagos, Value Analysis Consult. Ahmed R.B (2000),

Accounting Theory, 4th Edition, U.K, Business Press Thomson Learning.
Akanni J.A. (1988), Management, concepts, techniques and cases, Ibadan, Julab Publications Ltd. American Accounting Association (AAA), (1966), A Statement of Basic Accounting Theory (ASOBAT), USA Research Bulletins.
American Accounting Association (AAA), (1965), Concepts and Standards Research Study Committee. The Realisation Concept, the Accounting Review.Pp206 -210.
American Accounting Association, (AAA),(1964) the motivational Assumptions for Accounting Theory, the Accounting Review, Vol. 39, No. 3, July.
American Institute of Certified Public Accountants, Practical Guide on Accounting for uncertain Tax Positions under Fin 48, Nov. 29, 2006.
American Institute of Certified Public Accountants, (AICPA), (1970) Basic Concepts and Accounting principles underlying Financial Statement of Business Enterprises, Statement No. 4, New York, Accounting principles Board (APB)
American Accounting Association (AAA), (1936) Concepts and Standards of Accounting Theory, Accounting Review, Vol. 33, June. Pp216-318.
Anao A.R. (1996), An Introduction to Financial Accounting, Lagos, Longman Nigeria Ltd.
Baxter, W.T. (1981), Depreciation: Depreciating Assets - An Introduction, USA, Gee & Co.
Benjamin, O. (1990), Studies in Accountancy, Text and Readings, Enugu, New Age Publishers.
Bierman, H. and Drebin, A.R. (1972), Financial Accounting, An Introduction, Collier, Macmillan.
British Institute of Management Information, (1984), The Accounting Period, Note, 22.
Buckmaster O and Brooks L.D., Effect of price Level Changes in Operating Income CPA Journal May, 1974.
Burk, M. (1973), Consistency and Accounting changes, Texas, CPA.
Chambers R.J. (1966), Accounting Evolution and Economic Behaviour, USA, Prentice- Hall.
Chase, K.W. The Limits of Materiality, C.A. Magazine (Canada), June, 1979 Chatfield, Micheal (1977), A History of Accounting Thought New York, Kreiger
Dopuch, N. and Sunder, S. FASB's Statements on Objectives and Elements of Financial Accountings; a review (1980), The Accounting review, Vol. LV, No. 1, January.
Edey, H.E. (1970), The Nature of profit, Accounting and Business Research, No. 1 winter.
Fisher, I. (1980), Stabilising the Dollar, USA, Macmillan.
Fremgen, J.M. (1968) "The Going Concern Assumption" A Critical Appraisal, The Accounting Review, October. Vol43 (4).Pp649-656.
Fu P (1971), Government Accounting in China, during the Chou Dynasty (1122BC - 256 BC), Journal of Accounting Research. Vol 9(1).Pp40-51.
Gulman, S. (1939), The Accounting Concepts of Profits, New York, Ronald Press.
Glautier M.W.E and Underdown B. (2001) Accounting Theory and Practice, 7th Ed., New York, Prentice Hall Financial Times.
Glenn, L.J. and James A.G (1974), Finney and Miller's Principles of Accounting: Intermediate, 7th Edition, New Jersey, Prentice-Hall Inc.
Goldberg (1949), The Development of Accounting in Gibson, Accounting Concepts Reading, Cassell.
Grady, P. (1965), An Inventory of generally Accepted Accounting principles for Business Enterprises Accounting research Study Bo. 7, New York: AICPA.
Hendrikson E.S. (1992), Accounting Theory, 5 th Edition, USA, Richard D. Irwin.
Horngren, C.T. and George E. (1990), Cost Accounting: A Managerial Emphasis, 6th edition, New Delhi, Prentice Hall of India private Ltd.
Husband, GR., (1954) "the Entity Concept in Accounting: The Accounting Review, October, Vol. 29(4) Pp552-563.
Hylton, D.P. (1965), Matching Revenues with Expenses, The Accounting review, New York City.
Institute of Chartered Accountant in England and Wales, (1975), Recommendation on Accounting Principles
International Financial Reporting Standards - IFRS, (2007). The Full Text of International Financial Reporting Standards extant at 1 January 2007, International Accounting Standards Board Committee Foundation (IASCF), London, United Kingdom.
International Accounting Standards Committee - IASC (2001). The Full Text of International accounting Standards and SIC Interpretations, London, IASB Publications.
James M.E. (1955), Estate Account of the East of Northumberland (1562-16370, London, Surtes Society.
Jennings A.R. (1990), Financial Accounting, London, DP. Publications.
Keister O.R (1965), The Mechanics of Mesopotamian Record-keeping. The National Association of Accountants Bulletin. II.
Keynes L. (1980), "In The Long Run we are all dead" the Penguin Dictionary of Modern Quotation by J.M. and M.J. Cohen, Penguin.

Littleton, A.C. (1966), The Functional Development of Double-Entry, The Accountant, New York, Text Press.

Macre, R. (1981), a Conceptual Framework for Financial Accounting and Reporting, Accounting standard Committee

Members of ICAEW, The development of Accounting Study in UK dates from 1970.

Mike, H. and Fred K. (1983), Financial Accounting Theory and Standards, 2nd Edition, Great Britain, Prentice Hall International, Inc.

Millichamp A.N. (1990) Auditing: An Instructional Manual for Accounting Students, 5th Edition, London, DP Publication Ltd.

Moore, M.L. (1972), Conservatism, Texas, CPA.

Mootze M, (1970), Three contributions to the development of Accounting principles prior to 1930, Journal of Accounting Research, AAA. Vol. 8(1) Pp145-155.

Mootze, M. (1983), The development of Accounting theory, 3rd Edition, New York, Texas Inc.

Nwoko Chinedu (1990), Studies in Accounting, Text and Reading, Enugu, New Age Publishers.

Okoye I.C (2003), Research Manual, Guide for Research in Applied Science, Education, Technology, Medicine, Engineering and Business, Yola Paraclete Publishers.

Okoye, A.E. (1997), Cost Accountancy: Managerial Operational Applications, Benin, United City Press.

Ola C.S., (1985), Accountancy for Higher Exams, Ibadan, University Press Ltd. Omolehinwa Eddy (2004), Coping with Cost Accounting, Lagos, Pumark Ltd.

Omolehinwa, O. (2003), Foundation of Accounting, Lagos Pumark Nigeria Ltd.

Osuala E.C. (2005) Introduction to Research Methodology, 3rd Edition, Enugu, Africana First Publishers Limited.

Paton and Littlelon (1940), An Introduction to Corporate Accounting Standards, AAA.

Paton W.A. (1922), Accounting Theory, reprinted 1973, Lawrence, Kansas, Scholars Books Co. and 1996, Matthew and Parera, Cataloguing-in-Publication Data.

Paul Gee (1985), Book-keeping and Accounts, 19th Edition, London, Butterworth & Co Ltd.

Perara M.H.B and Matthew MR (1996), Accounting theory and Development Australia, International Thomson Publishing Company.

Pyle W. White J. and Zin M. (1980), Fundamental Accounting Principles, 3rd Edition, USA, Richard D. Irwin Inc.

Pyle, White, Larson and Zin (1988), Fundamental Accounting Principles, 3rd edition, USA, Richard D. Irwin Inc.

Remi, A., (2006), Advanced Financial Accounting, Second Edition, Lagos Master Stroke Consulting.

Richard J.B. (1981), Introduction to Accountancy and Finance, London. The Macmillan Press Ltd.

Robert O.I. (1999), Financial accounting Made Simple, Lagos, ROI Publishers.

Robert, G.M. Joesl, E.R. and James, R.C. (1990), Information Systems for Modern Management, 3rd Edition, New Delhi, Prentice Hall of India Ltd.

Rorem C.R. (1937) Accounting Theory: A critique of the Tentative Statement of Accounting principles, Accounting Review. Vol.12 (2) Pp133-138.

SSAP 4, The Accounting Treatment of Government Grant, ASC, April, 1974.

Sterling R.R. (1972), An Explication and Analysis of the Structure of Accounting, USA, Abacus.

Stoner J, Freeman R. and Gilbert, D Jr. (2002), Management, Sixth Edition, India, Prentice Hall.

Turpin, P.H and Stein, N.D (1986), Q & A on Financial Accounting Advanced Techniques, 2nd Edition, Great Britain, Financial Training Publications Limited.

Wolk Harry I, Dodd James L and Rozycki John J (2008). Accounting Theory: Conceptual Issues in a Political and Economic Environment, 7th edition, Sage Publications Inc. California.

Yamey, B.S. (1940). The Functional Development of Double-entry, The Accountant, New York, Text Press. Vol. 103Pp333-342

Yamey B.S. (1980), early views on the origins and Development on Book-Keeping and Accounting, Accounting and Business Research (Special Accounting History) issues, New York, Standard Text Press. Pp81-91